\documentstyle[12pt,epsfig]{article}
\newcommand{\spur}[1]{\not\! #1 \,}   
\begin{document}
\date{}
\title{\small \phantom{.}\hfill BARI-TH 320/98 \\[1cm]
\large\bf DYNAMICAL GENERATION OF THE \\
PRIMORDIAL MAGNETIC  FIELD BY\\  
FERROMAGNETIC DOMAIN WALLS}
\author{Paolo Cea$^{a,b}$ and Luigi Tedesco$^{b}$} 
\maketitle
\begin{center}
$^{a}$Dipartimento di Fisica and $^{b}$INFN, Sezione di Bari \\
Via Amendola 173,  I-70126 Bari, Italy*
\end{center}
\vspace{2cm}
\begin{abstract}
The spontaneous generation of uniform magnetic condensate in $QED_3$ gives rise 
to ferromagnetic domain walls at the electroweak phase transition. These 
ferromagnetic  domain walls are caracterized by vanishing effective surface 
energy density avoiding, thus, the domain wall problem. Moreover we find that 
the domain walls generate a magnetic field
$B \simeq 10^{24} Gauss$ 
at the electroweak scale which 
account for the seed field in the so called dynamo mechanism for the 
cosmological primordial magnetic field. 
We find that the annihilation processes 
of walls with size $R \simeq 10^5 Km$ could release an energy of order $10^{52} 
erg$ indicating  the invisible ferromagnetic walls as  possible compact 
sources of Gamma Ray Bursts.
\end{abstract}
\newpage
Recently there was a renewed interest in the phenomenon of the spontaneous 
generation of the magnetic condensate both at zero~\cite{hosotani}
and finite~\cite{das,ceat} temperature in three-dimensional 
spinorial electrodynamics.
The interest in three-dimensional QED is not merely academic. Indeed, it is 
well known that Dirac fermions with a Yukawa coupling with a scalar field 
develop zero mode solutions near a domain wall~\cite{jackiw}. These zero modes 
behave like massless fermions in the two-dimensional space of the wall. It has 
been suggested that ferromagnetic domain walls colud be relevant for the 
formation of the primordial magnetic field~\cite{iwazaki}. The aim of this 
paper is to discuss a dynamical mechanism which induces an uniform magnetic 
condensate localized on the wall. Moreover, as we shall discuss, we find a 
natural solution  of the domain wall problem so that we feel that our proposal 
provides a viable and realistic mechanism to account for the origin of the 
primordial magnetic field. \\
The only elementary scalar fields known in high energy physics are given by the 
scalar sector of Standard Model. According to the Standard Model when the 
Universe temperature reaches  the electroweak scale $T_c \simeq 10^2 Gev$, 
there is a phase transition where the scalar 
field aquires a vacuum expectation  
value. 
In order to avoid inessential compliances we shall consider a 
simplified model which, however, retains the essential aspects of the full 
electroweak theory. In our model, the scalar sector is given by a real scalar 
field with Lagrangian:
\begin{equation}
\label{eq1}
{\cal{L}}_{\phi}= \frac {1} {2} ( \partial_{\mu} \phi)^2 - \frac {\lambda} {4}
(\phi^2 - v^2)^2 .
\end{equation}
At the phase transition the vacuum aspectation value of the scalar field 
assumes the values  $\langle \phi \rangle = \pm v.$ So that one may assume 
that there are regions with $\langle \phi \rangle = + v$ and 
$\langle \phi \rangle = - v$. It is easy to see that the classical equation of 
motion admits the solution describing the transition layer between two regions 
with different values of $\langle \phi \rangle$:
\begin{equation}
\label{eq2}
\phi(z)= v \tanh \left(\frac {z} {\Delta}\right),
\end{equation}
where
\begin {equation}
\label{eq3}
\Delta= \frac {1} {v \sqrt{\frac {\lambda} {2}}} 
\end{equation}
is the thickness of the wall. As it is well known there is some additional 
energy associated with the wall proportional to the area. The surface energy 
density is given by:
\begin{equation}
\label{eq4}
\sigma_{\phi}= \int dz \left[ \frac {1} {2} (\vec{\nabla} \phi)^2 +
\frac {\lambda} {4} (\phi^2 - v^2)^2\right] = \frac {2 \sqrt{2}} {3} 
 v^3 \sqrt{\lambda}.
\end{equation}
Let us consider, now, a four dimensional massless Dirac fermion coupled to the 
scalar field throught the Yukawa coupling:
\begin{equation}
\label{eq5}
{\cal{L}}_{\psi} = \bar{\psi}(x)  i \spur \partial \psi(x) - g_{Y} \bar{\psi}(x) 
\psi(x) \phi(x).
\end{equation}
In presence of the kink Eq.(\ref {eq2}) the Dirac equation
\begin{equation}
\label{eq6}
[i \spur \partial - g_Y \phi(x)] \psi(x)=0
\end{equation}
admits the zero energy solutions localized on the wall \cite{jackiw}. Using the 
following representation of the Dirac algebra:
\begin{equation}
\label{eq7}
{\gamma}^3= i \left(
\begin{array} {cc}
0 & 1 \\
1 & 0
\end{array}
\right), \; \;
{\gamma}^{\alpha}= i \left(
\begin{array} {cc}
{\tilde{\gamma}}^{\alpha} & 0 \\
0  & -{\tilde{\gamma}}^{\alpha}
\end{array}
\right), \; \; 
\alpha=0,1,2
\end{equation}
with
\begin{equation}
\label{eq8}
{\tilde{\gamma}}^0=\sigma_3, \; \; 
{\tilde{\gamma}}^1=i \sigma_1, \; \; {\tilde{\gamma}}^2=i \sigma_2,
\end{equation}
we find for the zero modes:
\begin{equation}
\label{eq9}
\psi_0(x,y,z) = \frac {N} {\sqrt{2}} \; \omega(z) \left(
\begin{array} {cc}
\rho(x,y) \\
\rho(x,y)
\end{array}
\right)
\end{equation}
\begin{equation}
\label{eq10}
\omega(z)= exp\left\{- g_Y \int_0^z \phi(z') d z' \right\}
\end{equation}
where  $\rho(x,y)$ is a Pauli spinor, and 
\begin{equation}
\label{eq11}
N=\left[\int^{+\infty}_{-\infty} d z \; \omega^2 (z) \right]^{- \frac {1} {2}}=
\sqrt{\pi} \Delta \frac {\Gamma (g_Y \Delta)} {\Gamma (g_Y \Delta+ \frac {1} 
{2})}.
\end{equation}
It is easy to see that $\rho(x,y)$ satisfies the $2+1$-dimensional massless 
Dirac equation with the Dirac algebra Eq. (\ref {eq8}). \\
After the electroweak 
spontaneous symmetry breaking the unique long range gauge field is the familiar 
electromagnetic field $A_{\mu}$. In our scheme the real scalar field is the 
physical Higgs field, so that $A_{\mu}(x)$ does not couple to $\phi(x)$. Thus 
the relevant Lagrangian turns out to be:
\begin{equation}
\label{eq12}
{\cal {L}} = - \frac {1} {4} F_{\mu \nu}(x) F^{\mu \nu} (x) + \bar {\psi} (x)
[i \spur \partial - e \spur A  (x)] \psi(x) - g_Y \bar {\psi} (x) \psi (x) \phi 
(x).
\end{equation}
We are interested in the case of an external magnetic field localized on the 
wall at $z=0$ and directed along the $z$-direction: 
\begin{equation}
\label{eq13}
A_{\mu}( \vec {x}) = \left(0, \frac {B} {2} y, - \frac {B} {2} x, 0 \right).
\end{equation}
Moreover we shall restrict our analisys to the low-lying modes localized on the 
wall. In these approximations it is straightforward to check that the 
Hamiltonian reduces to:
\begin{equation}
\label{eq14}
H=\int d^3 x \frac {B^2} {2} + \int d^2 x \; \rho^{\dag}(x,y) H^{(2)} \rho(x,y)
\end{equation}
where
\begin{equation}
\label{eq15}
H^{(2)}= {\tilde {\alpha}}^k \cdot (i \partial_k - e A_k)
\end{equation}
and ${\tilde {\alpha}}^k = {\tilde {\gamma}}^0 {\tilde {\gamma}}^k, k=1,2.$
If we take into account the finite width of the wall $\Delta$ we may assume 
that the magnetic field $B$ is sizeable over a distance $z \simeq \Delta$.
So that Eq.(\ref {eq14}) can be rewritten as 
\begin{equation}
\label{eq16}
H= A \Delta  \frac {B^2} {2} + \int d^2 x \; \rho^{\dag}(x,y) H^2 \rho(x,y),
\end{equation}
where $A$ is the area of the wall. In the following we shall consider the 
surface energy density. Equations (\ref {eq15}) and (\ref {eq16}) show that the 
zero mode contribution to the energy is given by two-dimensional massless Dirac 
fermions. Note that the fermionic contribution to $H$ does not depend on the 
width of the wall.  In the case of two-dimensional massive Dirac fermions it 
is known since long time that it is energetically favorable the spontaneous 
generation of an uniform constant magnetic field~\cite{cea}. Indeed the 
fermionic energy density is  given by~\cite{cea}:
\begin{equation}
\label{eq17}
{\cal{E}}_F(B)=\frac {|eB|^{\frac {3} {2}}} {2 \pi} g(\xi), \; \;  
\xi = \frac {|eB|} {m^2}.
\end{equation}
Near the origin one finds:
\begin{equation}
\label{eq18}
{\cal {E}}_F(B)= - \frac {|eB|} {4 \pi} |m| + \frac {1} {24 {\pi}^2} 
\frac {(eB)^2} {|m|}
\end{equation}
The negative linerar term in Eq. (\ref{eq18}) gives rise to the negative 
minimum in the vacuum energy density. It is remarkable  that this phenomenon 
survives to the thermal corrections, even in the infinite temperature 
limit~\cite{ceat}. 
\\
On the other hand it can be shown that in $2+1$-dimensional $QED$ in presence 
of an external omogeneous background magnetic field with massless Dirac 
fermions it is energetically favourable to generate a constant mass term.
To see this we follow the Cornwall, Jackiw and Tomboulis formalism of the 
effective potential for composite operators~\cite{cjt}. If we restrict to the 
case of a constant Dirac mass term coupled with the local operator $\bar {\rho} 
(x) \rho(x)$, $\rho (x)$ being the three-dimensional Dirac field, then the 
relevant quantity turns out to be:
\begin{equation}
\label{eq19}
\Omega= {\cal{E}} (B, m) - {\cal {E}} (B, 0)
\end{equation}
where ${\cal{E}}(B,m)$ and ${\cal{E}}(B,0)$ are respectively the expectation value 
of the Hamiltonian Eq. (\ref{eq15}) on the massive and massless ground state.
After some calculations, which will be reported elsewhere, we find in the 
one-loop approximation:
\begin{eqnarray}
\label{eq20}
\Omega(B,m) &=& - \frac {|eB|} {2 \pi} 
\sum_{n=1}^{\infty} \sqrt{2 n |eB| + m^2} +
m^2 \frac {|eB|} {2 \pi} \sum_{n=1}^{\infty} \frac {1} 
{\sqrt{2 n |eB| + m^2}} + \nonumber \\
&&
- \frac {\sqrt{2} \zeta (\frac {3} {2})} {8 \pi^2} 
|eB|^{\frac {3} {2}}.
\end{eqnarray}
A dimensional analysis shows that $\Omega(B,m)=|m|^3 f(\xi)$, where we recall 
that $\xi=\frac {|eB|} {m^2}$. In Figure 1 we report $f(\xi)$ versus $|m|$ for 
fixed $|eB|.$ We see clearly that the massive ground state lies below the 
massless ground state  for any value of the mass. Moreover the minimum is 
obtained 
for infinite mass. The above analysis strongly suggests that in three 
dimensional $QED$ with a massless Dirac fermion there is spontaneous generation 
of both a mass term for the 
fermion field and a uniform magnetic field.  In  this case,
from Eqs. (\ref{eq16},\ref{eq18}) we have for the surface energy density of the 
ground state 
\footnote{Strictly speaking Eq. (\ref{eq21}) is valid
for $\frac {|eB|}
{m^2}<<1$. As we shall see, our results show that indeed this is the physically 
relevant region.}
\begin{equation}
\label{eq21}
\sigma_{QED}= {\cal {E}}_{QED} (B,m)= \Delta \frac {B^2} {2} - \frac {|eB|} {4 
\pi} |m| + \frac {1} {24 \pi^2} \frac {|eB|^2} {|m|}.
\end{equation}
Remarkably Eq. (\ref{eq21}) display a negative minimum at 
\begin{equation}
\label{eq22}
B^*=\frac {e|m|^2} {4 \pi} \frac {1} 
{|m| \Delta +\frac {e^2} {12 \pi}}
\end{equation}
with
\begin{equation}
\label{eq23}
\sigma_{QED}= {\cal {E}}_{QED}(B^*,m)=
- \frac {e^2} {32 \pi^2} 
\frac {|m|^3} 
{|m| \Delta +\frac {e^2} {12 \pi}}.
\end{equation}
To summarize, we have found that in presence of the kink solution Eq. (\ref 
{eq2}) of the Higgs field there are massless fermion modes localized on the 
wall of the kink. In the one-loop approximation the zero modes becoming massive 
generate a constant magnetic field on the wall lowering the surface energy 
density by the amount given by Eq.(\ref {eq23}). Thus the total
 surface energy 
density of the wall is:
\begin{equation}
\label{eq24}
\sigma=\sigma_{\phi}+\sigma_{QED} = \frac {2 \sqrt {2}} {3} v^3 \sqrt {\lambda} 
- \frac {e^2} {32 \pi^2} 
\frac {|m|^3} 
{|m| \Delta +\frac {e^2} {12 \pi}}.
\end{equation}
where we recall that $\Delta= \frac {\sqrt {\lambda}} 
{v \sqrt{2}}.$
\\
The negative surface energy density induced by the magnetic condensation
increases with the fermion zero mode mass. However the stability of the whole 
system requires that $\sigma \geq 0$. As a consequences the strength of the 
magnetic field increases up to the value such that 
\begin{equation}
\label{eq25}
\sigma=0.
\end{equation}
By using the standard values $v\simeq 250 \; Gev$,   
$\alpha_{QED}= \frac {e^2} {4 
\pi} \simeq \frac {1} {137}$ and assuming $\lambda \simeq 0.5$ 
we find that  Eq.(\ref {eq25}) admits one and only one 
positive root for $|m|$:
\begin{equation}
\label{eq26}
|m|\simeq 1.7 \cdot 10^4 \;  Gev,
\end{equation}
and 
\begin{equation}
\label{eq27}
B^* \simeq 5 \cdot 10^4 Gev^2 \simeq 5 \cdot 10^{24} \; Gauss.
\end{equation}
Note that $ \frac {e B^*} {m^2} <<1$ so that the approximation 
implied in Eq. (\ref {eq21}) is rather good.
\\
It is worthwhile to stress the main results we obtained. The remarkable 
phenomenon of spontaneous generation of an 
uniform magnetic condensate in 
$(2+1)$-dimensional $QED$ gives rise to ferromagnetic domain walls which are 
invisible as concern the gravitational interactions. Indeed, according to 
Eq.(\ref 
{eq25}) the surface energy density  of the ferromagnetic walls is zero, 
thus solving the gravitational domain wall problem. In addition our invisible 
domain walls generate a magnetic field at the electroweak scale 
whose strength Eq. (\ref {eq27}) agrees remarkably well with the 
one needed for the initial seed field in the dynamo mechanism for generating
the cosmological magnetic field~\cite{tornk}. Finally our proposal for the 
invisible ferromagnetic domain walls could be relevant for the problem of the 
Gamma Ray Bursts~\cite{piran}. 
Indeed the collision of  a kink with an 
anti-kink should proceed like the annihilation of a fermion-antifermion pair.
In the annihilation processes the negative energy of the magnetic condensate 
becomes disposable for the production of gamma particles. The energy for these 
processes can be estimated from Eq. (\ref{eq23}). Indeed, 
from Equations (\ref{eq26})
and (\ref{eq27}) we get 
\begin{equation}
\label{eq28}
\sigma_{QED} \simeq -4.2 \cdot 10^7 {Gev}^3,
\end{equation}
which corresponds to the huge amount of disposable energy per ${fm}^2$
\begin{equation}
\label{eq29}
E \simeq 1 {fm}^2 \cdot |\sigma_{QED}| \simeq 10^9 \; Gev.
\end{equation}
If we require that the released energy in such process is of 
order $10^{52} erg$,
then according to Eq. (\ref{eq29}) we find that the 
linear dimension of the wall 
should be about $10^5 Km$. So that we see that our  proposal provides 
relatively compact sources to account for the Gamma Ray  Burst.
\\


%
\newcommand{\InsertFigure}[2]{\newpage\phantom{.}\vspace*{-3cm}%
\phantom{.} \vspace*{5.truecm}\hspace*{.5truecm}
\vspace*{-8.truecm}
\begin{center}\mbox{%
\epsfig{bbllx=8.truecm,bblly=2.truecm,bburx=16.5truecm,bbury=28.truecm,%
height=18.truecm,figure=#1}}
\end{center}\vspace*{-.1truecm}%
\parbox[t]{\hsize}{\small\baselineskip=0.5truecm\hskip0.5truecm #2}}

\InsertFigure{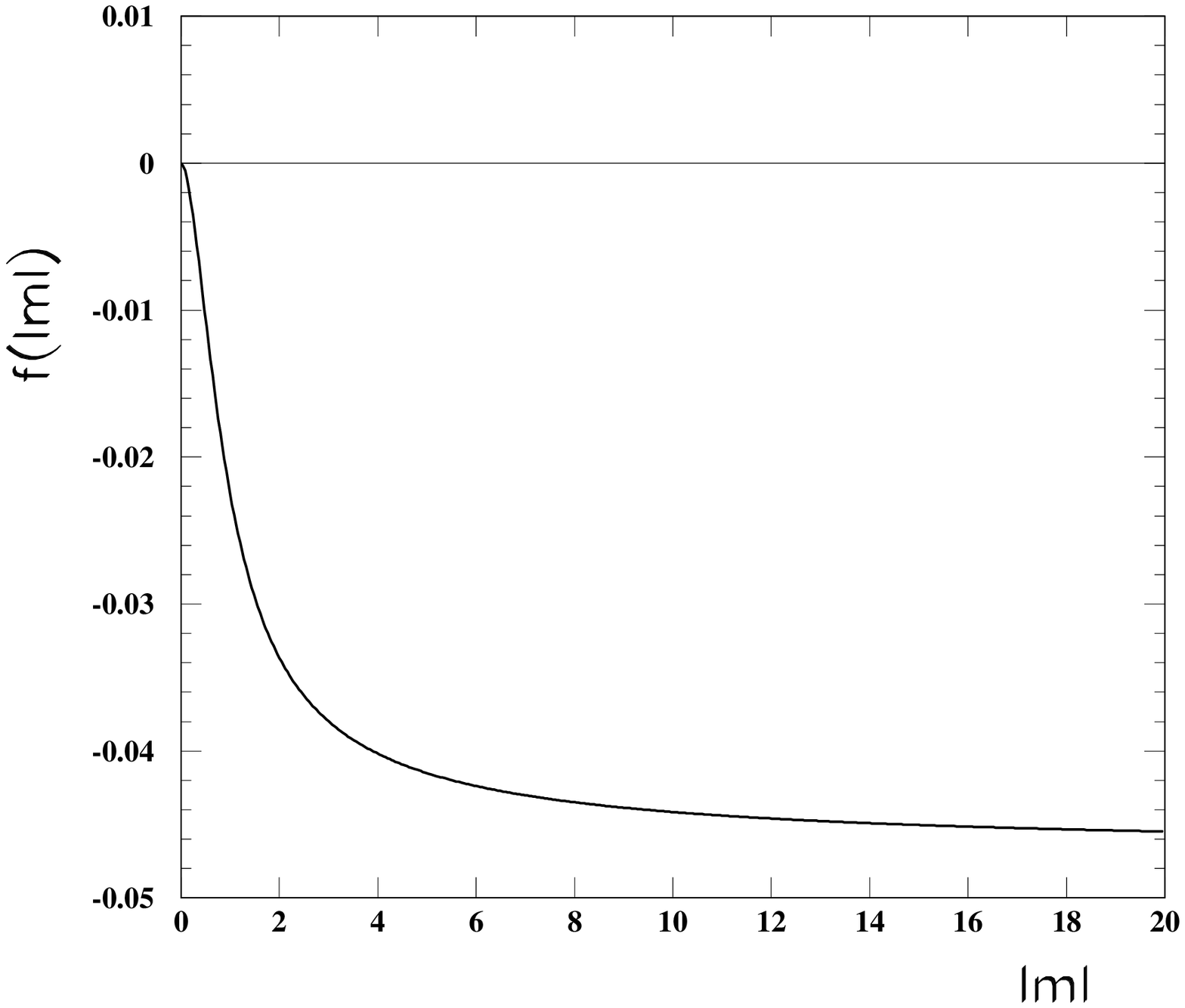}%
{                   The one-loop effective potential $\frac {\Omega(B,m)}
                   {|m|^3}$ versus $|m|$ for $|eB|=1$.}


\newpage

\begin{thebibliography}{99}
%
\bibitem{hosotani}     Y. Hosotani, Phys. Lett. B {\bf 319} (1993) 332;
                       Phys. Rev. D {\bf 51} (1995) 2022;
                       D. Cangemi, E. D'Hoker, and G. Dunne, Phys. Rev. 
                       D {\bf 51} (1995) 2513;
                       V.P. Gusynin, U.A. Miransky, and I.A. Shovkovy, Phys. 
                       Rev. Lett {\bf 73} (1994) 3499;   {\bf 76} (1996) 1005; 
                       Nucl. Phys. B {462} (1996) 249;
                       V. Zeitlin, Phys. Lett. B {\bf 352} (1995) 422;
                       P. Cea, Phys. Rev.  D {\bf 55} (1997) 7895.
%
\bibitem{das}          A. Das and M. Hott, Phys. Rev. D {\bf 53} (1996) 2252;
                       S. Kanemura and T. Matsushita, Phys. Rev. D {\bf 56} 
                       (1997) 1021.
%
\bibitem{ceat}         P. Cea and L. Tedesco, Phys. Lett. B {\bf 425}
                       (1998) 345.
%
\bibitem{jackiw}       R. Jackiw and C. Rebbi, Phys. Rev. D {\bf 13} (1976)
                       3398; A.T. Niemi and G.W. Semenoff, Phys. Rep. C {\bf 
                       135} (1986) 100.
%
\bibitem{iwazaki}      A. Iwazaki, Phys. Lett B {\bf 406} (1997) 304; 
                       Phys. Rev. D {\bf 56} 1997 2435.
%
\bibitem{cea}          P. Cea, Phys Rev. D {\bf 32} (1985) 2785; Phys. Rev. D 
                       {\bf 34} (1986) 3229.
%
\bibitem{cjt}          J.  Cornwall, R. Jackiw and E. Tomboulis, Phys. Rev. D 
                       {\bf 10} (1974) 2428.
%
\bibitem{tornk}        For a review see O. Tornkvist, {\it 
                       Magnetic Fields from The 
                       electroweak Phase Transition}, hep-ph/9801286; 
                       Phys. Rev. D {\bf 58} (1998) and references therein.
%
\bibitem{piran}        For a recent review see T. Piran, {\it 
                       Gamma-Ray Burst and 
                       Related Phenomena}, astro-ph/9801001 and references 
                       therein.



\end{thebibliography}
\end{document}